\documentclass{jetpl}
\twocolumn

\lat
\usepackage{amsmath}
\usepackage{amsfonts}
\usepackage{amssymb}
\usepackage{amstext}
\usepackage{graphicx}
\usepackage{color}
\graphicspath{{images/}}

\title{Dynamic susceptibility of a spin ice near the critical point}

\rtitle{Dynamic susceptibility of a spin-ice\ldots}

\sodtitle{Dynamic susceptibility of a spin-ice near the critical point}

\author{A.\,V.\,Shtyk$^{a,b}$ and M. V. Feigel'man$^{a,b}$}

\rauthor{A.\,V.\,Shtyk, M. V. Feigel'man}

\sodauthor{Shtyk, Feigelman}
\address{$^a$ L. D. Landau Institute for Theoretical Physics, Kosygin str.2, Moscow
119334, Russia,$^b$ Moscow Institute of Physics and Technology, Moscow 141700, Russia}

\abstract{
We consider spin ice magnets (primarily, $\mathrm{Dy_2Ti_2O_7}$) in
the vicinity of their critical point on the $(H,T)$ plane. We
 find that the longitudinal susceptibility diverges at the critical
 point, leading to the behaviour qualitatively similar to the one
which would  result from non-zero conductance of magnetic charges.
We show that dynamics of  critical fluctuations  belongs to  the universality
class of  easy-axis ferroelectric
 and calculate logarithmic corrections (within two-loop approximation)
 to the mean-field critical behavior.
}


\begin{document}

\maketitle

\begin{figure}[tbp]
\includegraphics[width=1\linewidth]{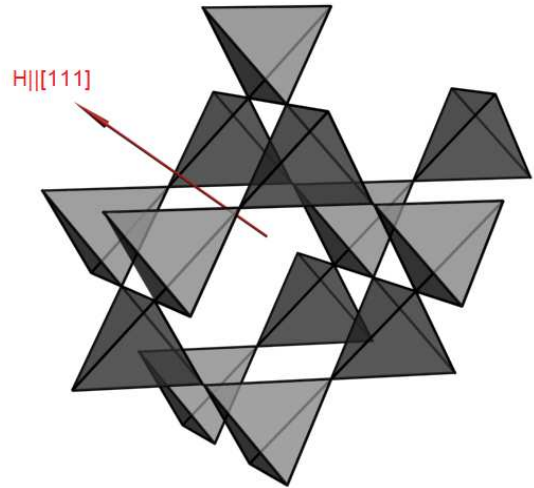}
\caption{Fig.1.  Pyrochlore lattice consists of corner-sharing tetrahedra;
spins of dysprosium are siting at their vertices.}
\label{Pyrochlore}
\end{figure}

Spin ice~\cite{Gingras} is a kind of geometrically frustrated
magnetic materials with a structure of low-energy states  presenting
deep similarity with usual water ice, those H-bond structure was analized long
ago by Pauling~\cite{Pauling}. The most studied examples of spin ice
include pyrochlore  oxides $\mathrm{Dy_2Ti_2O_7}$ and
$\mathrm{Ho_2Ti_2O_7}$; their structure is shown in Fig.~\ref{Pyrochlore}.
 Spin ice demonstrates a number
 of peculiar low-temperature properties, the major of the are: i) an extensive residual entropy,
 and ii) elementary excitations  resembling magnetic
 monopoles\cite{CMS,Gingras}.
 Low temperature dynamics of spin ice is governed almost solely by
 these magnetic excitations, therefore one may expect
  spin ice to demonstrate  phenomena similar to
  those known for the electrolytes. Indeed, it was
  recently shown that some aspects of spin ice behaviour  can be
  described in terms of  ``magnetolyte'' similar to the
  Onsager theory~\cite{Onsager}  of electolytes~\cite{magnetolyte}.
  It is also known~\cite{Gingras},  that at low enough temperatures spin ice
  undergoes a first-order transition as function of applied magnetic
  field $H$. The first-order transition line in the
  $(H,T)$ plane terminates at the critical point $H_c,T_c$.
  In particular, $H_c= 0.9 \mathrm{T}$ and $T_c = 0.38 \mathrm{K}$ in the case of  $\mathrm{Dy_2Ti_2O_7}$.
  In the present Letter we calculate dynamic longitudinal magnetic susceptibility of
  spin ice near this critical point; our results predict  the
  behaviour formally similar to the one expected for the media with a
  nonzero ``magnetoconductance''. We will explain, however, that it
  would be incorrect to  interpret these results in terms of ``direct
  current'' of monopoles. Specific numerical estimates will be done
  for $\mathrm{Dy_2Ti_2O_7}$, whereas our general scheme is applicable
 to  any pyrochlore spin ice.

The crucial difference between  spin ice and electrolyte is the finite
static magnetic susceptibility of the former as opposed to the diverging
dielectric low-frequency response $\epsilon(\omega) \propto i\sigma/\omega$
 of any electrolyte with a nonzero conductivity $\sigma$. The origin
 of this difference can traced to the fact that magnetic monopoles of
 spin ice  are the sources of magnetization field $\mathbf{M}$ \, (or,
 equivalently,  magnetic field $\mathbf{H}$), but they \textit{do not generate}
magnetic induction $\mathbf{B}$,  which is the proper analog of electric field $\mathbf{E}$
for electrostatics~\cite{LandauLifshits9}.

 Following~\cite{Ryzhkin}, one can represent linear magnetic
 susceptibility of spin ice as
\begin{equation}
\chi(\omega,T)=\frac{1}{\chi_0^{-1}(T)-i\omega/\Gamma}\label{susceptibility_simple}
\end{equation}
where $\chi_0(T) \sim \mu_B^2/(V_0T)$ is  static susceptibility (
$\mu_B$ is the Bohr magneton and $V_0$ is the volume of an elementary cell).
 Therefore  relaxation of magnetization after the sudden switching
 of magnetic field is given by
\begin{equation}
M(t)=\chi_0H(1-\exp[-\Gamma \chi_0^{-1}t])
\label{Mt}
\end{equation}
 At the small timescales $t\ll \chi_0/\Gamma$  the relaxation process (\ref{Mt}) resembles
direct conductivity with "magnetic current" proportional to the external magnetic field:
 $j_m = dM/dt = \Gamma H$, thus the magnetic analog of conductivity is $\sigma_m = \Gamma$.
However, finite value of static susceptibility $\chi_0(T)$ limits possible observation of
magnetoconductivity to a transient regime  only.

We  emphasize that Eq.(\ref{susceptibility_simple}) for dynamic susceptibility $\chi_0(H,T)$ with some effective
relaxation rate  $\Gamma(H,T)$ should be valid in a broad range of temperatures and fields, when monopoles are present.
Spin ice in the absence of external field, and spin ice in sufficiently strong field when kagome ice structure
develops, represent the two well-defined limiting cases.
Below we consider the vicinity of the critical point $(H_c,T_c)$ where static susceptibility is
expected to diverge, leading to a broad timescale for the observation of magnetoconductivity.
We will use virial expansion in order to estimate the parameters
of the Ginzburg-Landau free energy  $F\{m\}$ which describes fluctuations in the critical region.
Then we use this $F\{m\}$ functional to  construct an effective  Martin-Siggia-Rose functional~
\cite{MSR,Hohenberg,Lebedev} which describes critical dynamics of spin ice,
and study it up to the two-loop logarithmic approximation.
\begin{figure}[tbp]
\includegraphics[width=1\linewidth]{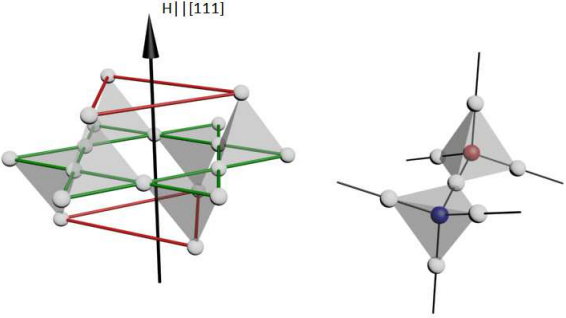}
\caption{Fig.2. Pyrochlore lattice can be represented as a stack of triangular and kagome lattices, with easy axis directions being aligned with the bonds of diamond lattice, which is dual to the pyrochlore lattice.
Magnetic monopoles are situated in the centers of tetrahedra.}
\label{Pyrochlore_2}
\end{figure}

The  Hamiltonian of the spin ice is of  the form\cite{Gingras}
\begin{eqnarray}
\label{hamiltonian}
H=-\frac{1}{2}\sum\limits_{i, j}J_{ij}(\mathbf{e}_i\cdot \mathbf{e}_j)S_iS_j - g\mu_B \mathbf{H} \sum_i \mathbf{e}_iS_i \\ \nonumber
+ Da^3\sum\limits_{(ij)}\left[\frac{\mathbf{e}_i\cdot \mathbf{e}_j}{|\mathbf{r}_{ij}|^3}-
\frac{3(\mathbf{e}_i\cdot \mathbf{r}_{ij})(\mathbf{e}_j\cdot \mathbf{r}_{ij})}{|\mathbf{r}_{ij}|^5}\right]S_iS_j \notag
\end{eqnarray}
Here $S_i = \pm 1$ is an effective Ising variable which describes sign of  magnetic moment
$\mathbf{\mu}_i = S_i \mathbf{e_i}$ of the $i$-th Dysprosium ion; the magnitudes
$|\mathbf{\mu_i}|=\mu \approx 10 \mu_B$,
whereas $\mathbf{e}_i$ are the unit vectors along easy axis, see Fig.~\ref{Pyrochlore_2}, the direction of axes being chosen with respect to one of the tetrahedral sublattices, see Fig.~\ref{Pyrochlore}.
 Parameters  $J_{ij}$ describe exchange interactions between first-, second- and third-order neighbours,
their magnitudes were determined~\cite{Tabata}
 as  $J_1=-3.72\mathrm{K}$, $J_2=0.1\mathrm{K}$ and $J_3=-0.03\mathrm{K}$ respectively.
External magnetic
field $\mathbf{H}$ is directed along the $[111]$ axis,
and  the last term in Eq.(\ref{hamiltonian}) stands for the magnetic dipole-dipole interaction,
the dipole interaction constant $D = \mu^2/a^3 = 1.41 \mathrm{K}$, where $a = 3.54\text{\AA} $
is the pyrochlore lattice constant (it is equal to the nearest-neighbour distance, see Fig.~\ref{Pyrochlore_2}).

We will study the system described by the Hamiltonian (\ref{hamiltonian}) near the critical point,
 where large
 average magnetization $\bar{M}_\parallel$ along the field direction is induced. Deviation $m$ of the
actual magnetization $M_\parallel$
from its average value at the critical point $\bar{M}_{\parallel}(H_c,T_c)$ can be considered as an
order parameter which describes the state of the system in  the vicinity of the critical point.
The corresponding Ginzburg-Landau free energy functional is of the form
\begin{equation}
\mathcal{F}=\int dV \left[\frac{1}{2}m(a+\hat{b})m+\frac{\lambda}{4!}m^4-mh\right]
\label{F}
\end{equation}
where $a=\alpha(T-T_c)$ and  $h=H-\gamma(T-T_c)$. In a system with local interactions
one would find $\hat{b}=bk^2=-b\Delta$, but our case is different due to the presence of  strong dipole-dipole interactions.
Similar problem of  second-order phase transition in uniaxial ferroelectric was considered in the seminal paper~\cite{LK},
where it was shown that in presence of dipole-dipole interaction operator $\hat{b}$ should be modified as follows:
\begin{equation}
\hat{b}=bk^2+4\pi x^2
\end{equation}
where $x=k_z/k$ and the z is [111] axis.
We are interested here in the frequency-dependent susceptibility of spin ice,
and thus we need to extend the renormalization group analysis
developed  in Ref.~\cite{LK} for the critical dynamics.
Due to absence of any locally conserved quantities in the problem, the system can be described by purely
 relaxational dynamics governed by the free energy (\ref{F}):
\begin{equation}
\partial_t m=-\Gamma \frac{\delta\mathcal{F}}{\delta m}
\label{dynamics}
\end{equation}
To obtain estimates for coefficients in (\ref{F}) we use virial expansion as described below; the estimate
for the kinetic coefficient $\Gamma$ will be provided at the end of this Letter.
At low temperatures $T \ll T_c$, in the high-$H$ phase  almost all spins are aligned with the magnetic field,
 while in the low-$H$ phase the same is valid for the  spins of triangular sublattices only;  configurations
 of other spins is governed by the kagome ice rules, see Fig. 2 and Ref.\cite{Gingras}.  One possible way to
 estimate the parameters
of $\mathcal{F}$ near the critical point would be to consider gas of interacting monopoles taking into
 account their
direct as well as entropic interactions~\cite{MS}; similar problem of lattice ion systems was considered
 in~\cite{LS}. However, we prefer to start from the high-$H$ phase and to consider an exponentially dilute
(far from the critical point) gas of flipped spins as a lattice gas of interacting particles.
Such an approach will be approved \textit{a posteriori}
by the numerically small concentration of flipped spins even at the critical point.
 Using the Hamiltonian
(\ref{hamiltonian}) we obtain the
 energy of such a particle (siting in the site $0$) in the form
\begin{eqnarray}
E_0=\frac{2}{3}\mu H+2\sum\limits_{i}J_{0i}(\mathbf{e}_0\cdot \mathbf{e}_i)S_i- \\
2Da^3\sum\limits_{i}\left[\frac{\mathbf{e}_0\cdot \mathbf{e}_i}{|\mathbf{r}_{0i}|^3}-\frac{3(\mathbf{e}_0\cdot \mathbf{r}_{0i})(\mathbf{e}_i\cdot \mathbf{r}_{0i})}{|\mathbf{r}_{0i}|^5}\right]S_i\notag
\end{eqnarray}
where $S_i=1$ on triangular sublattices and $S_i=-1$ on kagome sublattices.
The interaction of two particles siting in the sites $i$ and $j$ is equal to
\begin{equation}
U_{ij}=\frac{4}{3}J_{ij}+4Da^3\left[\frac{\mathbf{e}_i\cdot \mathbf{e}_j}{|\mathbf{r}_{ij}|^3}-\frac{3(\mathbf{e}_i\cdot \mathbf{r}_{ij})(\mathbf{e}_j\cdot \mathbf{r}_{ij})}{|\mathbf{r}_{ij}|^5}\right]
\end{equation}
Now we can employ virial expantion~\cite{Kubo} in terms of small density of particles $n$, to obtain
 free energy
\begin{equation}
F(n)=TN_0[-1-n-b_2n^2-(b_3-2b_2^2)n^3+\frac{E_0}{T}n+n\ln n]
\end{equation}
where $T$-dependent functions
\begin{eqnarray}
\label{b2b3}
b_2=\frac{1}{2!}\sum\limits_{i}f_{0i},\quad\text{with}\quad
 f_{ij}=\exp\left[-\frac{U_{ij}}{T}\right]-1\notag
\\ \nonumber
b_3=\frac{1}{3!}\sum\limits_i\sum\limits_j[f_{0i}f_{ij}f_{j0}+f_{0i}f_{0j}+f_{i0}f_{ij}+f_{0i}f_{0j}]\notag
\end{eqnarray}
are "cluster integrals" given by the summation over the lattice.
We calculate these sums numerically and  find the position of the critical point
$H_c^{th}=1.37 T$, $T_c^{th}=0.85 K$ and critical concentration of ``particles'' $n_c = 0.14$
from the conditions
\begin{equation}
\frac{dF}{dn}=0,\quad\frac{d^2F}{dn^2}=0,\quad\frac{d^3F}{dn^3}=0
\end{equation}

Smallness of $n_c$ supports qualitative validity of our virial expansion; however, experimental
 values of $H_c$ and $T_c$ are below our estimates by factors 1.5-2. With the position of the critical
point being determined, we can  estimate values of the parameters entering free energy functional
$\mathcal{F}$:
\begin{equation}
\alpha=\frac{da}{dT}=50\mathrm{K}^{-1},\quad\lambda=1300\text{\AA}^{3}\cdot\mathrm{K}^{-1},\quad\gamma=0.2\mathrm{T}\cdot\mathrm{K}^{-1}\label{estimates}
\end{equation}
However, virial expansion is not useful to determine the coefficient $b$ entering the gradient term in (\ref{F});
thus we estimate it within nearest-neighbours approximation, neglecting long-distance tail of the dipole-dipole
interaction:
\begin{equation}
b\approx\frac{5\sqrt{2}}{24}\frac{J_1+5D}{D}a^{2}\approx26\text{\AA}^{2}
\end{equation}

Coming back to the problem of longitudinal susceptibility and considering it in the mean-field approximation,
 one obtains (1) with $\chi_0^{-1}=\alpha(T-T_c)$,
so exactly in the critical point $\chi\propto{i}\Gamma/\omega$.
 We will see below that account of fluctuations do not change this result considerably,
leading to logarithmic
 corrections only.

\begin{figure}[tbp]
\includegraphics[width=1\linewidth]{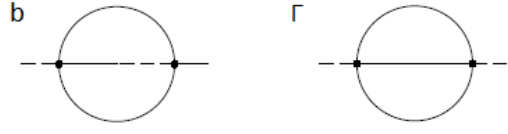}
\caption{Fig. 3. Diagrams responsible for the renormalization of coefficients $b$ and $\Gamma$.
Solid lines represent correlation function $\langle m_1m_2 \rangle$, and mixed lines correspond to
 response function  $\langle m_1p_2 \rangle$.}
\label{diagrams}
\end{figure}

With the parameters of the free energy (\ref{F}) specified, we
move to the analysis  of critical fluctuations and their role in the dynamics.
Thus we construct, using Eqs.(\ref{F},\ref{dynamics}), the  Martin-Siggia-Rose~\cite{MSR,Hohenberg,Lebedev}
 dynamic action
\begin{equation}
\mathcal{I}=\int dtdV[\Gamma^{-1} p\partial_tm+apm+p\hat{b}m+\frac{\lambda}{6}pm^3-ph+iT\Gamma^{-1} p^2]
\label{MSR}
\end{equation}
where $p=p(t,\mathbf{r})$ is the dynamic field conjugated to the magnetization field $m=m(t,\mathbf{r})$,
 and the last term in the action
describes thermal noise. The estimate for the kinetic coefficient $\Gamma$ will be provided later.
Summation of leading logarithmic corrections with the action (\ref{MSR}) up to one-loop approximation
can be done in complete analogy with the paper of Larkin and Khmelnitsky~\cite{LK}, as it refers to the
 static quantities only:
\begin{equation}
\frac{d\lambda}{d\xi}=-g\lambda,\quad\frac{da}{d\xi}=-\frac{1}{3}ga
\end{equation}
\begin{equation}
g=\frac{3T}{16\pi\sqrt{4\pi b^3}}\lambda,\quad\xi=\frac{1}{2}\ln\frac{b\Lambda^2}{bk^2+4\pi x^2}
\label{1loop}
\end{equation}
where $\Lambda\sim a^{-1}$ is an ultraviolet cutoff.
Renormalization of the gradient term $b$ and the kinetic coefficient $\Gamma$ appears in the second-loop approximation only.
The corresponding diagrams are represented in Fig.\ref{diagrams}. Calculation
 of these diagrams leads to the  renormalization-group equations
\begin{eqnarray}
\label{2loop}
-\frac{d\Gamma}{d\xi}=\frac{2C}{3^3\pi^6}g^2\Gamma,\, \frac{db}{d\xi}=\frac{2^2}{3^5}g^2b,\, C=\int\limits_0^\infty\frac{dpds}{p^3s^2}f^3(s,p)\\
f(s,p)=\int\limits_{q>0}\frac{q^2dqd\xi d\varphi}{q^2+\xi^2}\exp[-\frac{q^2+\xi^2}{s} +i\frac{q}{p}\cos\varphi+iq\xi]  \notag
\end{eqnarray}
where the constant $C\approx 400$. The solution of equations (\ref{2loop}) is as follows:
\begin{eqnarray}
g=\frac{g_0}{1+g_0\xi},\quad
a=\frac{a_0}{(1+g_0\xi)^\frac{1}{3}}\\
b=b_0\exp[\frac{4}{3^5}\frac{g_0^2}{1+g_0\xi}]  \notag\\
\Gamma=\Gamma_0\exp\left[-\frac{2C}{3^3\pi^6} \frac{g_0^2}{1+g_0\xi}\right]  \notag\\
\xi=\frac{1}{2}\ln\frac{b\Lambda^2}{|a|+bk^2+4\pi x^2+|\omega|/\Gamma}  \notag
\end{eqnarray}
Using Eqs.(\ref{estimates},\ref{1loop}) we find  that initially value of coupling constant is $g_0=0.15 \ll 1$.
 Then  renormalization of $b$ and $\Gamma$ is very  small numerically  and we neglect it, so the susceptibility takes the form
\begin{equation}
\chi=\frac{1}{a(k,x,\omega)+bk^2+4\pi x^2-i\omega/\Gamma}
\end{equation}
and in the uniform field
\begin{equation}
\chi=\frac{1}{a-i\omega/\Gamma},\quad a(T)=\frac{\alpha(T-T_c)}{(1+\frac{1}{2}g_0\ln\frac{b\Lambda^2}{\alpha|T-T_c|+|\omega|/\Gamma})^\frac{1}{3}}
\end{equation}
For $T<T_c$ static susceptibility should be modified by a factor of two: $\alpha$ should be substituted by $\alpha/2$. This is the result in the absence of effective field $h$ which is defined after Eq.(\ref{F}).
If the temperature $T=T_c$,  then the renormalization cutoff will be determined by $h$, i.e. by the deviation $H-H_c$
\begin{equation}
\chi=\frac{1}{a^*-i\omega/\Gamma},\quad a^*(h)=\frac{\left(\frac{9}{2}\lambda_0h^2\right)^{\frac{1}{3}}}{(1+\frac{1}{2}g_0\ln\frac{b\Lambda^2}{(\lambda_0h^2)^{\frac{1}{3}}+ |\omega|/\Gamma})^{\frac{1}{3}}}
\end{equation}

Exactly at the critical point $a=0$ and $\chi \sim i\omega^{-1}$.
However, in this system there is no conductivity in the usual sense. The point is that our lattice
consists of a stack of interchanging kagome and triangular layers. In moderate magnetic fields (in particular,
 in the vicinity  of the critical point), orientation of spins on the triangular layers are almost fixed.
 This results in the situation when
monopoles are bound to theirs kagome layers and thus there is no direct motion of
 monopoles in the direction
 of the field. In fact, monopoles of one kagome layer reside in centers of tetrahedra, i. e. on
 two sublattices that are  separated by a finite distance $h=a_d/3$ in the direction $(1,1,1)$
of the applied field.
 Thus positive-charge and negative-charge monopoles will be siting, preferably, on the "upper" and
"lower" sublattices correspondingly, with the energy difference $\mu H/3$ between them.
As a result, magnetization is primarily determined by the concentration of the monopoles.

Finally, we need to estimate relaxation rate $\Gamma$.  Dynamical processes in spin ice at low
 temperatures and in low-field phase are governed almost solely by monopoles. Ground state doublet of each
 spin is separated by large energy gap $\Delta\sim300\mathrm{K}$ from higher-energy states(\cite{Gingras}).
 Thus, at temperatures $T \leq 1 \mathrm{K}$, processes of spins flips are solely quantum.
 Then,  the dependence
 of magnetoconductivity on temperature follows Arrhenius law with activation energy $E_m \approx 1 \mathrm{K}$
 corresponding to the creation  of one monopole. Reminding the relation between $\Gamma$ and $\sigma$, we find
\begin{equation}
\Gamma(T,H=0)=\Gamma_0 e^{-E_m/T}
\label{Gamma1}
\end{equation}
where $\Gamma_0$ is temperature-independent.
Low-field dynamical magnetic susceptibility of $\mathrm{Dy_2Ti_2O_7}$ was measured in Ref.~\cite{Susceptibility}.
 Using the results~\cite{Susceptibility} for the imaginary part of susceptibility together
with Eq.(\ref{susceptibility_simple}), we obtain
\begin{equation}
\Gamma(T=1.8\mathrm{K},H=0)\sim 100 s^{-1}\nonumber
\end{equation}
Comparing it with Eq.(\ref{Gamma1}) and the estimate for $E_m$.
 we conclude that $\Gamma_0 \sim 100 s^{-1} $ as well.

The nature of spin relaxation in spin ice near critical point is not quite clear.
As we already mentioned above, in presence of magnetic field $\sim 1 T$,
 positive and negative magnetic charges reside
 on different sublattices and the tunneling processes become essentially modified, as
the positions of monopoles on the kagome lattices are not equivalent. There are
two possible scenarios for tunneling processes: sequential  and simultaneous flips of two spins.
Below we estimate the rate of sequential process, which will provide the lower bound for the
relaxation rate near critical point.
Energy difference for the monopoles of the same sign siting in two kagome sublattices
leads to the  exponential ratio of their concentration:
$N_2=N_1\exp[-2\mu H/3T]$. On the other hand, detailed balance condition reads
$N_1 \Gamma_1=N_2\Gamma_2$
where subscript  ``1'' refers to monopoles situated on "native" sublattice, and
subscript  ``2'' to "foreign" ones. Assuming $\Gamma_2 \sim \Gamma_0$ since no
additional barrier exists for the hopping  of monopole from a ``foreign'' site to a ``native one'',
 we come to the estimate
\begin{equation}
\Gamma^{upper}(H,T)\sim\Gamma_0(0,T)\exp[-2\mu H/3T]
\label{Gamma_upper}
\end{equation}
and eventually $\Gamma_c\sim(0.01\div 100)s^{-1}$.

In conclusions,  we calculated dynamic spin susceptibility of spin ice material
$\mathrm{Dy_2Ti_2O_7}$ near the critical point. Logarithmic corrections due to critical
fluctuations are found to be small. The response of the form
 $\chi(\omega) \sim (i\omega)^{-1}$ is predicted in the broad range of low frequencies.
Measurement of the prefactor in this dependence would allow to determine the nature
of elementary spin flip processes near critical point.

This research was supported by the RFBR grant \# 10-02-00554 and by the
RAS Program  ``Quantum physics of condensed matter'' . We are grateful to
S. Korshunov, I. V. Kolokolov and S. Sosin for useful discussions.

\end{document}